\documentclass[a4paper]{article}

\usepackage{INTERSPEECH2022}

\usepackage{booktabs}
\usepackage{multirow}
\usepackage{tabularx}
\usepackage{enumerate}
\usepackage{amssymb, arydshln, balance}
\usepackage{color}

\def\x{{\mathbf x}}

\usepackage{enumitem}

\title{Speaker conditioned acoustic modeling for \\ multi-speaker conversational ASR}

\name{Srikanth Raj Chetupalli and Sriram Ganapathy \thanks{This work was supported by the grants from  the British Telecom Research Center.}}
\address{LEAP lab, Electrical Engineering, Indian Institute of Science, Bangalore, India. 
\email{ \{sraj,sriramg\}@iisc.ac.in}}

\ninept
\begin{document}

\maketitle
\begin{abstract}
In this paper, we propose a novel approach for the transcription of speech conversations with natural speaker overlap, from single channel speech recordings. The proposed model is a combination of a speaker diarization system and a hybrid automatic speech recognition (ASR) system. The speaker conditioned acoustic model (SCAM) in the ASR system consists of a series of embedding layers which use the speaker activity inputs from the  diarization system to derive speaker specific  embeddings. The output of the SCAM are speaker specific senones that are used for decoding the transcripts for each speaker in the conversation. In this work, we experiment with the  automatic speaker activity decisions generated using an end-to-end speaker diarization system. A joint learning approach is also proposed where the diarization model and the ASR acoustic model are jointly optimized. The experiments are performed on the mixed-channel two speaker recordings from the Switchboard corpus of telephone conversations. In these experiments, we show that the proposed acoustic model, incorporating speaker activity decisions and joint optimization, improves significantly over the ASR system with explicit source filtering (relative improvements of $12\%$ in word error rate (WER) over the baseline system). 
\end{abstract}

\section{Introduction}
The transcription of single-channel natural long-form speech conversations is   desired for various applications like call center data, medical conversations, meeting data, court recordings, movie closed captioning, etc. Typically, the transcription of natural speech conversations involves the two processing steps of speaker diarization (SD)  and automatic speech recognition (ASR) that  are performed independently. The outputs of these models are then combined. However,  as previously noted by Shafey et. al. \cite{Shafey2019Joint},  such a processing pipeline is sub-optimal. 
In natural multi-talker conversations~\cite{watanabe2020chime6,singh2021leap}, the speech content is rich in speaker overlaps, back channels, and turn-taking.   This paper attempts to build a speech recognition system that uses time-varying speaker activity decisions in the acoustic model.

In the area of speaker diarization, the end-to-end neural diarization (EEND) approaches have  overcome some of the limitations of traditional systems (with i-vectors~\cite{zhu2016online} or x-vectors~\cite{singh2019leap}). Self-supervised learning has also been recently explored for diarization \cite{singh2021self, singh2021self2}.  The EEND models generate speaker activity predictions at each frame  \cite{fujita2019endtoend, fujita2019endtoend2, shinji2020Interspeech}. The architectures for EEND use bidirectional-LSTM (BLSTM) layers \cite{fujita2019endtoend2},  self-attentive (SA) transformer encoder layers \cite{fujita2019endtoend}, or  encoder-decoder attractor (EDA) layers   \cite{shinji2020Interspeech}. 

In the ASR literature, most of the works have focused on the single-talker speech in clean/noisy settings~\cite{agrawal2019modulation} or multi-talker speech segmented with reference speaker activity~\cite{saon2017english}. 
On the other hand, overlapped speech recognition has been explored primarily on artificial overlap generated by merging single talker speech recordings.
Several approaches to multi-talker ASR, based on source separation ~\cite{Yu2017Recognizing}, sequence transduction~\cite{seki2018purely} and end-to-end architectures were investigated recently \cite{8736286, neumann2020multitalker}.

\begin{figure*}[t]
    \centering
\includegraphics[width=17cm, height=6.0cm, trim={0.3cm 0.35cm 0.3cm 0.25cm}, clip]{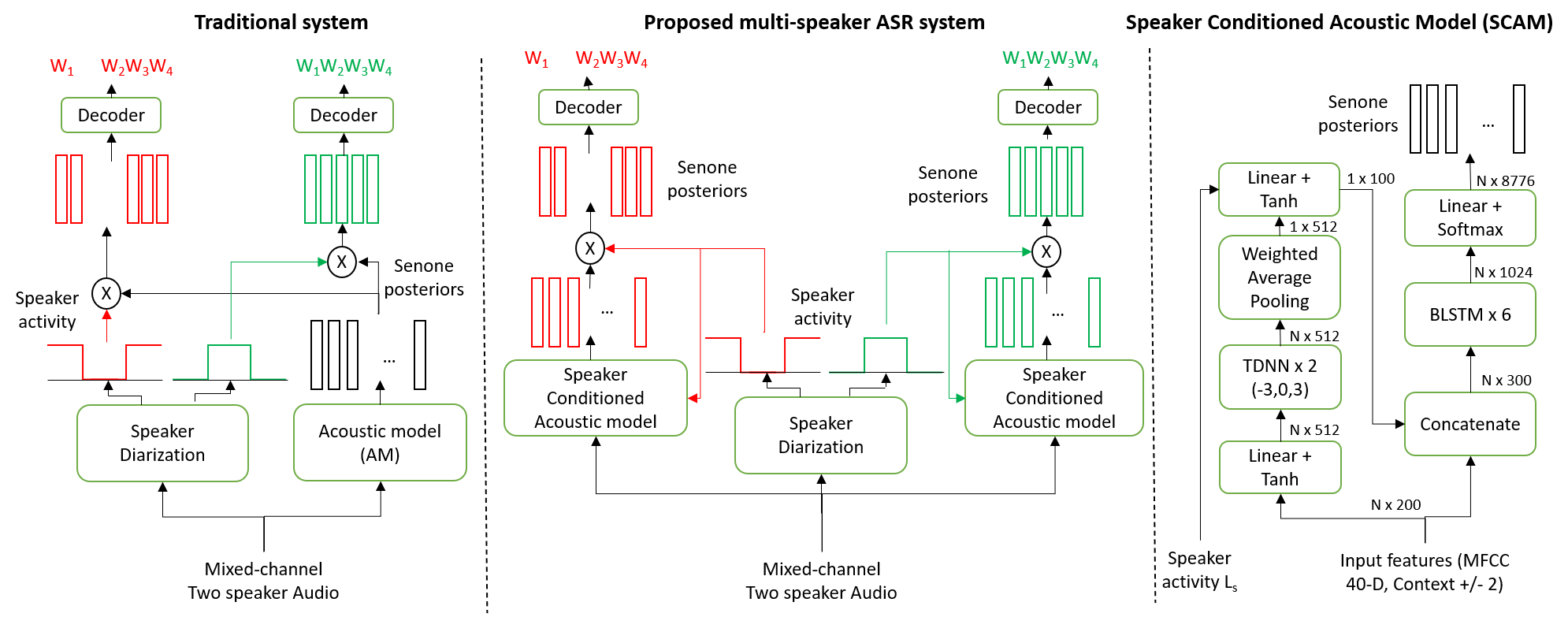}
\vspace{-0.3cm}
     \caption{Schematic of traditional ASR (left), proposed approach (center), and speaker conditioned acoustic model (SCAM) (right). } \label{fig:overall_blockdiagram}
     \vspace{-0.3cm} 
\end{figure*}


In this paper, we consider the transcription of multi-speaker speech conversations from a single-channel recording. The key novelty is the development of an acoustic model (AM) for a hybrid speech recognition system that is speaker aware. Using the input acoustic features and the speaker activity decisions from an external model, the acoustic model,  consisting of neural sub-networks, generates speaker specific embeddings internally. The speaker specific embeddings are used with the acoustic features for the prediction of speaker conditioned senones (context dependent hidden-Markov-model (HMM) states). 

The experiments are performed on the mixed channel conversational two-talker English telephone recordings from the Switchboard corpus~\cite{switchboard}. 
We observe that the baseline system, combining separate acoustic model and speaker diarization systems, degrades significantly in the mixed-channel recordings compared to the single-channel recordings. The proposed approach to AM training improves the ASR performance significantly (average relative improvement of $30$\%) over the competitive baseline. The joint training of the ASR and diarization systems further improves the ASR and achieves a performance similar to the system using single   speaker recordings.

\section{Related prior work}
 A single architecture for two-talker speech recognition was proposed in \cite{Yu2017Recognizing}, which was trained with a permutation invariant objective. For source separation based approaches \cite{8736286, neumann2020multitalker}, individual source signals/features are first obtained using a neural network model  \cite{kolbk2017multitalker, luo2017tasnet}, followed by a single channel ASR.   A sequence transduction approach is proposed in \cite{Shafey2019Joint} for joint speech recognition and speaker diarization. Here, the authors assume a non-overlapping speech scenario. The multi-speaker ASR for the End-to-End framework was explored in \cite{seki2018purely} and extended later in \cite{k2020investigation, k2020joint}. 
 
For the proposed acoustic model, the self-attentive end-to-end encoder-decoder-attractor (SA-EEND-EDA) model based neural diarization system of \cite{shinji2020Interspeech} is used to predict the frame-level speaker activity.  
The proposed approach  does not rely on explicit source separation, as done in \cite{8736286, neumann2020multitalker}. Also, the information about the participating speakers is not required, as in \cite{k2020joint}, and the model applies to overlapping speaker scenarios.

\section{Multi-speaker Conversational ASR}
Let $\boldsymbol{\mathcal{X}}$ denote the acoustic features of the speech input corresponding to a conversation. Here, $\boldsymbol{\mathcal{X}} = \{\x _1,... \x _N \}$, where $N$ is the number of time-frames and $\x$ is the $D$ dimensional acoustic feature. Let $S$ denote the number of speakers in the conversation.  In all the experiments reported in this work, $S=2$. Let $\boldsymbol{\mathcal{W}}^{(s)}$ be the set of words in the ground truth transcription spoken by $s^{th}$ speaker. 



\subsection{Speaker diarization}
We use the SA-EEND-EDA architecture proposed by Horiguchi et. al.  \cite{shinji2020Interspeech}.  Once the model is trained, at each time-frame, the network predicts the speech activity for each speaker. In this work, the SA-EEND model is specifically used for the two speaker case ($S=2$). The speaker activity outputs are binarized using a threshold of $0.6$ (chosen empirically based on the test set performance) and this output is denoted as $\boldsymbol{{L}}_s \in \boldsymbol{\mathcal{B}}^{N\times 1}$, for each speaker $s \in [1,\dots,S]$. More details about the model architecture  can be found in \cite{shinji2020Interspeech,fujita2019endtoend2}.
\subsection{Speaker conditions acoustic modeling}
 
The traditional ASR system (shown on the left of Figure~\ref{fig:overall_blockdiagram}) has the AM and the speaker diarization system operate independently. The speaker activity decisions from the diarization system are then combined with the AM outputs (senone posterior probabilities) to generate speaker specific outputs. These posterior probabilities are then fed to the decoder~\cite{mohri2002weighted}. 

The SCAM architecture (shown in the center of  Figure~\ref{fig:overall_blockdiagram}) consists of two branches, a speaker embedding generation branch and a senone posterior prediction branch, as shown in the right panel of Figure \ref{fig:overall_blockdiagram}. The $40$-dimensional mel-frequency cepstral coefficients (MFCCs), with a context of $\pm 2$ frames, form the acoustic features $(\boldsymbol{\mathcal{X}}) \in {\mathcal R}^{N \times 200}$. In the speaker embedding branch, these features are input to two time-delay neural network (TDNN) layers of output size $512$. The total context for the TDNN stack   is $\pm 8$ frames. The speaker activity decision from the diarization module ($\mathbf{L}_s$) for each speaker $s$ is used to perform a weighted pooling of the TDNN layer outputs. 

Specifically, the TDNN outputs are averaged over the given speaker's active time-frames to generate a pooled representation. The pooling layer output is further processed through a linear layer to a $100$-dimensional embedding space. The generated speaker-specific embedding ${\bf c}_s$ of dimension $100$ is concatenated with the input features $\boldsymbol{\mathcal{X}}$ and fed to a stack of $6$ BLSTM layers with $512$ units in each direction. The final BLSTM layer output is projected into the senone space ($8776$ dimensions) using a linear layer. 
The senone posteriors are computed independently for each speaker.
The final posterior probability matrix for each speaker $s$, $\mathcal{P}^{(s)}$, is then time segmented to the $N_s$ segments, 
and used in the decoder. This generates word sequences $\hat {\mathcal W}^{(s)}_{1:N_s}$ for  the $N_s$ speaker regions and for   the $S$ speakers. 

For the proposed SCAM model, even with the same acoustic speech features, the model is able to generate different senone posterior vectors based on the  speaker activity inputs. Further, the model is not constrained to the number of speakers in the input conversation as the model has the ability to  generate outputs specific to as many speakers hypothesized by the diarization system. The speaker activity detector from the diarization system, being a neural model  (SA-EEND-EDA system~\cite{shinji2020Interspeech}), integrates well with  the proposed neural SCAM model to form a full neural pipeline of processing with differentiable layers. In this manner, the speaker activity detector can be fine-tuned with the ASR loss function. 

\subsection{Baseline systems}

The architecture of the baseline model is similar to the right-side section in the SCAM model (Figure \ref{fig:overall_blockdiagram}) and it consists of a stack of $6$ BLSTM layers and an output linear layer. The network uses $40$-dimensional MFCC features with $\pm 2$ context and $100$-dimensional online i-vector features as the input \cite{garimella2015robust}. The online i-vectors replace the speaker activity based embeddings used in the proposed SCAM model. Thus, the number of parameters is identical to the SCAM. We consider two approaches, (i) training on clean, isolated channels (referred to as BLSTM-iso) and (ii) training on a single mixed channel (referred to as BLSTM-mix). During evaluation, mixed channel speech is input to the AM and single channel senone posteriors are generated by the AM. We also experiment with source separation (ConvTasNet~\cite{ConvTasNet}) followed by single channel ASR system on each of the source filtered outputs. 
\subsection{Performance metric}
In this work, we use a speaker specific word error rate (SWER) metric. 
For each speaker $s$, the word level transcription in the ground truth for the entire recording is concatenated and used as the reference. The model output in the form of concatenated word level transcripts from all the $N_s$ segments of the given speaker $s$ is used as the prediction. 
The WER for each speaker is measured and the aggregate SWER is then computed using all the speakers in the given recording. 
Thus, the SWER combines speaker segmentation errors and word transcription errors.
\subsection{Datasets and training}
The end-to-end speaker diarization model is trained on $100,000$ two-speaker mixtures, simulated using audio from Switchboard-2 \cite{switchboard2_phase1, switchboard2_phase2, switchboard2_phase3}, Switchboard Cellular \cite{switchboard_cellular1, switchboard_cellular2} and NIST-SRE (2004-08) datasets. The model is trained for $100$ epochs using binary cross-entropy loss with utterance-level permutation-invariant training (PIT). The default configurations from the reference implementation~\cite{fujita2019endtoend2}
are used for training the model. The model is further fine-tuned on the Switchboard-1 phase-III dataset \cite{switchboard}, for $100$ epochs. 
\begin{figure}[t!]
    \centering
    \includegraphics[width=2.5in,height=1.25in,trim={0 0 0 0},clip]{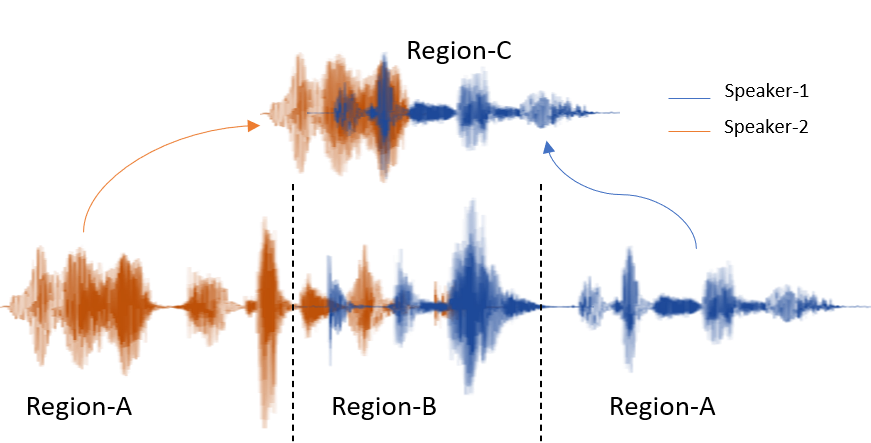}
    \caption{Segment definitions in mixed-channel speech. The region-A and region-B segments occur in natural conversations, while region-C is synthesized for data augmentation. }
    \vspace{-5pt}
    \label{fig:seg_illustration}
\end{figure}

We train the acoustic model on the $300$-hour Switchboard telephone conversations corpus \cite{switchboard}, and the tri-gram language model is trained on the Switchboard transcripts and Fisher English corpus transcripts. The Switchboard dataset consists of $2430$ two-sided telephonic conversations between $500$ different speakers and contains $3$M words of text. Each recording in the dataset consists of two channels, corresponding to the two (speaker) sides of the conversation. The two channels are mixed to form the single-channel in our training and test sets. The training dataset is also augmented with speed perturbed audio, with perturbation factors of $0.9$ and $1.1$. 

In the mixed channel speech (MCS), we define the segments as illustrated in Figure \ref{fig:seg_illustration}. The time-regions containing a single speaker are taken as-is and termed as region-A segments. We refer to the time-regions with natural overlap in the original recordings as region-B. During the ASR training, we augment the training set by artificially mixing segments of different speakers from region-A, as shown in Figure \ref{fig:seg_illustration}. For each segment of channel-1, a randomly selected channel-2 segment from the same conversation is mixed, creating an overlap. The fraction of overlap duration is chosen randomly in the range of $[30-70]$\%. We refer to the artificially created overlap segments as region-C segments. 

The training target for the AM contains two channels corresponding to the two speakers in the conversation. We obtain the speech features' alignment to the senones using a tri-phone model, with the recipe available in Kaldi~\cite{Kaldi}.
For region-A segments, one channel of the training target contains the senone indices obtained from alignment, and the senone label for the other channel is set to the index $0$. For region-B/C segments, the training target consists for the active speaker's  senone indices obtained from the ground truth alignment while the  label targets for the inactive  speaker are set to the index of $0$ (silence).  For region-C segments, the two-channel senone targets are composed using the senones of the individual segments used to create the overlap segments. The AM is trained using the cross-entropy loss with Adam optimizer and with a learning rate of $10^{-4}$. We use the PyTorch toolkit \cite{NEURIPS2019_9015} for the AM training, and the Kaldi toolkit for decoding. The training dataset statistics are given in Table \ref{tab:dataset_stats}. 
\begin{table}[t]
    \caption{Training dataset statistics}
    \vspace{-5pt}
    \centering
    \begin{tabular}{|p{4.5cm}|c|c|}
    \hline
    & Dataset & + Aug. \\
    \hline
    Total duration  & 751.6 hrs & 935.1 hrs\\
    \% overlap in duration  & 13.8  & 16.7\\
    \# segments & 259983 & 328929\\
    Duration of overlap segments  & 496.3 hrs & 679.7 hrs \\    
    \# overlap segments & 111,114 & 180,060 \\
    \% of overlap segments & 42.7 & 54.7 \\
    \% overlap duration in B,C segments & 20.9 & 24.1\\
    \hline
    \end{tabular}
    \label{tab:dataset_stats}
\end{table}

\begin{figure}[!t]
\vspace{7pt} 
    \centering
    \includegraphics[width=3in,height=0.9in,trim={15pt 32pt 40pt 60pt}]{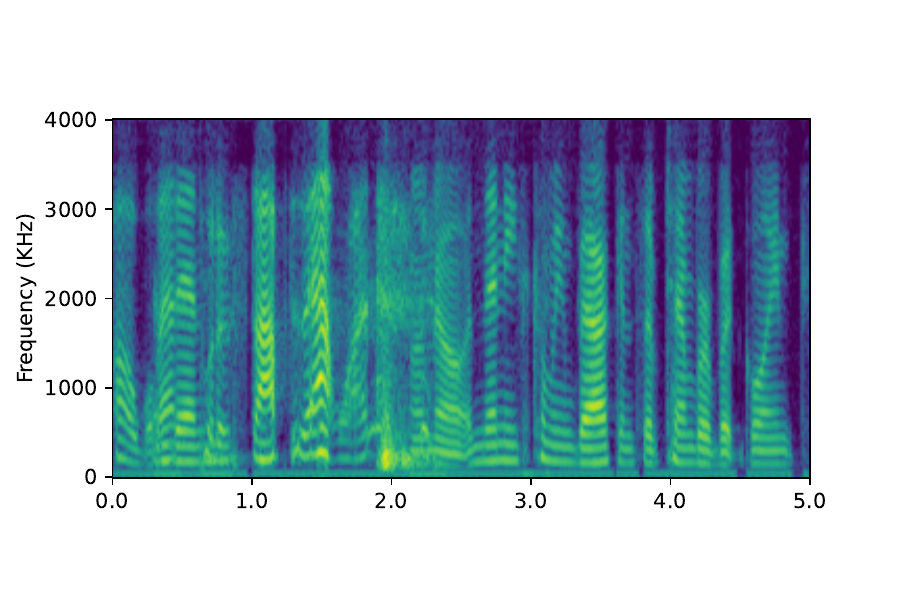}
    \includegraphics[width=3in,height=2.1in,trim={15pt 10pt 15pt 15pt}]{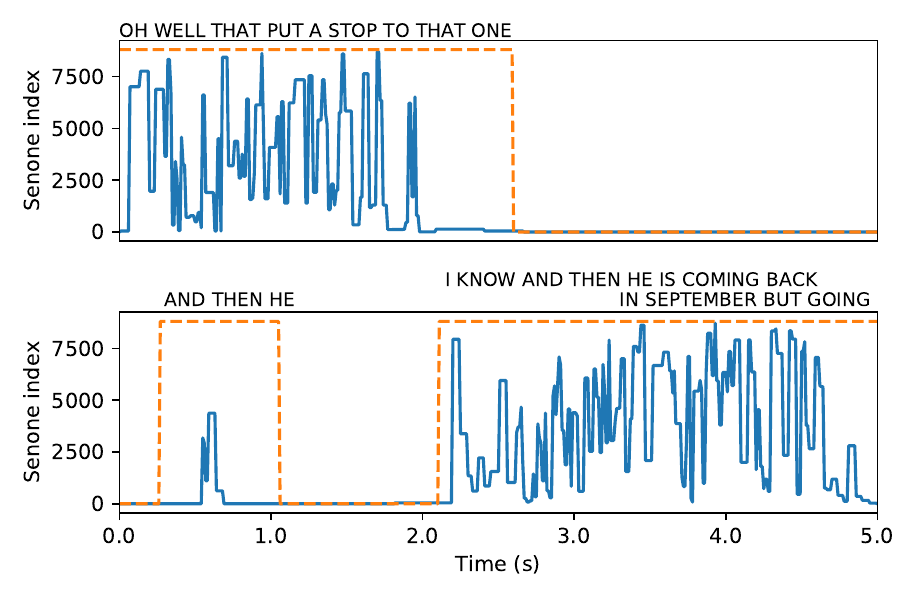}
     \vspace{-5pt} 
           \caption{Illustration of senone prediction using the SCAM model. The speaker activity is shown in dashed line.}
     \vspace{-5pt}
\label{fig:SCAM_illustrations}
\end{figure}

\begin{table}[t!]
\centering
\setlength{\tabcolsep}{1pt}
\renewcommand{\arraystretch}{0.95}
\caption{Example transcripts for the segment illustrated in Fig. \ref{fig:SCAM_illustrations}. The blue and red colors correspond to the two speakers. The ground truth transcript is shown in Figure~\ref{fig:SCAM_illustrations}.  }\label{tab:transcript_illustration}
\vspace{5pt}
\resizebox{1.0\columnwidth}{!}{
\begin{tabular}{l|l}
    
    \large{BLSTM-iso}& \large{SCAM} \\ \hline \\  
   \textcolor{blue} {OH WELL THAT AND WOULD'VE STOPPED } & \textcolor{blue}{     OH WELL THAT'S KIND OF STOPPED}\\
   \textcolor{blue}{TO THAT ONE [LAUGHTER] YOU KNOW} &  \textcolor{blue}{ TO THAT ONE [LAUGHTER]}\\ \\
  \textcolor{red}{WELL THAT AND WOULD HAVE }  & \textcolor{red}{AND THEN I KNOW AND THEN}     \\
  \textcolor{red}{YOU KNOW AND THEN HE'S COMING } & \textcolor{red}{HE'S COMING BACK IN SEPTEMBER } \\  
  \textcolor{red}{BACK IN SEPTEMBER BUT GOING} & \textcolor{red}{BUT GOING} \\ \hline
\end{tabular}}
\vspace{-10pt}
\end{table}

\section{Experimental setup} \label{sec:expt_results}
The evaluation data is the HUB5 English  speech (LDC2002S09 and LDC2002T43)  containing Switchboard subset similar to the training data. The dataset consists of $20$ telephone conversations from the Switchboard corpus and $20$ conversations from the CALLHOME American English speech corpus \cite{callhome}. We convert the two channel recordings into a single sum channel. The total duration of the evaluation dataset is $3.15$ hours, and $19.4\%$ of the total duration has speaker overlap. 

Figure \ref{fig:SCAM_illustrations} shows an illustration of the SCAM output for a short segment of speech. The ground truth speaker activity for the entire duration of $5$~s is given as input to the neural network. The plot shows the senone index, with maximum posterior probability, at each frame. During the overlap speech region, the network predicts different senones for the two speaker channels. The transcription obtained using the SCAM output is given in Table~\ref{tab:transcript_illustration}. 
The transcription obtained using the BLSTM-iso system (the baseline system) is also shown. The baseline  system makes more errors, and the transcription is similar for both the channels in the overlap region.



Table \ref{tab:asr_wer} shows the ASR WER performance for the systems compared. We evaluate the ASR in two ways, (i) with ground truth speaker activity (GTS), and (ii) with speaker activity obtained from the SA-EEND-EDA diarization. We train the SCAM model in four different ways, using all the segments with and without overlap speech augmentation, and, similarly, using segments with overlap speech only (either using region-B segments or using both region-B and region-C segments). The four versions are referred to as SCAM-[V1 - V4] in Table \ref{tab:asr_wer}. 

We experiment with computing the embedding (weighted average pooling) over speech frames where only the current speaker is active during evaluation (excluding overlap speech regions); we refer to the embeddings generated in this scheme as ``clean'' embeddings. The SCAM model inference performed using the clean embeddings is referred to as the V5 setting in Table~\ref{tab:asr_wer}. 
We also experiment with the joint training of the speaker activity detector and the SCAM modules in the proposed system. The trained SCAM-V4 model and the pre-trained SA-EEND-EDA models are fine-tuned for two epochs with the acoustic model loss as the criterion for optimization. The corresponding result (Joint training) is shown in Table \ref{tab:asr_wer}.

The first row in Table \ref{tab:asr_wer} shows the WER when the individual channels (not mixed) of the conversations are input to the BLSTM-iso system (baseline). \textcolor{black}{We also compare the performance of the proposed approach using explicit source separation followed by ASR. The pre-trained ConvTasNet model \cite{ConvTasNet}, available in \cite{Pariente2020Asteroid} is used to for source separation. The fourth row in Table \ref{tab:asr_wer} shows the WER when the separated channels are input to the BLSTM-iso system}. 

\begin{table}[t]
\caption{Speaker specific WER (SWER) \% for the baseline BLSTM systems and the speaker conditioned AM based systems.}
\vspace{-5pt}
    \centering
    
    \begin{tabular}{|p{0.8cm}|c|c|c|c|c|c|}
    \hline
    \multicolumn{1}{|c|}{\multirow{2}{*} {System}} & \multicolumn{2}{|c|}{Data set} & Aug. & \multirow{2}{*}{GTS}  & \multirow{2}{*}{Diar.}\\ \cline{2-4}
    & Reg.-A & Reg.-B & Reg.-C  &   & \\ 
    \hline
     \multicolumn{6}{|c|}{Single-channel} \\ \hline 
    \multicolumn{1}{|c|}{BLSTM-iso} & $\checkmark$ & $\times$ & $\times$   & 21.4  & 24.0  \\   
     \hline      
    \multicolumn{6}{|c|}{Mixed-channel} \\ \hline 
    \multicolumn{1}{|c|}{BLSTM-iso} &  $\checkmark$ & $\times$ & $\times$  & 37.7  & 37.2  \\
    \multicolumn{1}{|c|}{BLSTM-mix} & $\checkmark$ & $\checkmark$ & $\times$  & 40.5  & 40.6  \\  \multicolumn{1}{|c|}{ConvTasNet} & $\checkmark$ & $\times$ & $\times$   & 25.3  & 27.4  \\
    \hdashline  
    \multicolumn{1}{|c|}{SCAM-V1} & $\checkmark$ & $\checkmark$ & $\times$ & 28.3  & 27.9 \\
    \multicolumn{1}{|c|}{SCAM-V2}& $\checkmark$ & $\checkmark$ & $\checkmark$ & 28.2  & 27.7 \\
    \multicolumn{1}{|c|}{SCAM-V3}  &$\times$  & $\checkmark$ & $\times$ & 28.8 & 28.6\\
    \multicolumn{1}{|c|}{SCAM-V4}& $\times$ & $\checkmark$& $\checkmark$ &  27.5 &  26.9\\ 
     \multicolumn{1}{|c|}{SCAM-V5} & $\times$& $\checkmark$ & $\checkmark$   & 26.0 &  26.1\\ \hdashline  
    \multicolumn{1}{|c|}{Joint training}& $\times$& $\checkmark$ & $\checkmark$   & \textbf{23.4} & \textbf{24.1}\\ 
    \hline      
    \end{tabular}
    \vspace{-10pt}   
    \label{tab:asr_wer}
\end{table}

\begin{table}[t]
\caption{Speaker specific WER (SWER) \% for different test segment types: the single speaker (region-A) and overlap regions (region-B).  }
\vspace{5pt}
    \centering
    \begin{tabular}{|p{4cm}|c|c|}
    \hline
    \multicolumn{1}{|c|}{System} & Region-A & Region-B\\ 
    \hline
    \multicolumn{1}{|c|}{BLSTM-iso}       & 27.1 & 41.5\\
    \multicolumn{1}{|c|}{BLSTM-mix}       & 30.1 & 44.3\\ 
    \multicolumn{1}{|c|}{SCAM-V4} & 31.4 & 26.1\\
    \multicolumn{1}{|c|}{SCAM-V5} & 29.8 & 24.7\\    
    \hline      
    \end{tabular}
    \vspace{-10pt}
      
    \label{tab:asr_wer_rA}
\end{table}
The mixing of the speaker channels degrades the WER by an absolute margin of $16\%$ for the GTS scenario. The proposed SCAM models show significant improvements over the baseline BLSTM models. We see that the performance is better for the SCAM trained using the augmented dataset. Without augmentation, training on the whole dataset has better performance than training on the overlap (region-B) segments alone. However, when augmentation is used, we see that the SCAM trained only on  the overlap segments (region-{B,C}) alone is moderately better. Finally, the use of pure speaker embeddings improves further over the model V4. 

Table \ref{tab:asr_wer} also shows that the WER for the system  with the predicted speaker activity is similar to the system using the ground truth segmentation information for models without joint learning. Further, the joint learning  approach yields an absolute improvement of $13.1$ \% in WER over the baseline system with automatic diarization (relative improvement of $35$ \% over the baseline system). 

Comparing the source separation with ASR (ConvTasNet~\cite{ConvTasNet}), the proposed joint training approach improves significantly (relative improvements of $12$\%). As seen here, the results from the proposed joint modeling approach on the mixed channel speech gives ASR performance close to the single channel result with diarization outputs. The source separation approach is also more computationally involved compared to the proposed framework.  Further, unlike the source separation based approaches, the proposed framework is not restricted to two speaker conversations and can be used for recordings with arbitrary number of speakers. 


\par Next, we study the WER of the system for the single speaker (region-A) segments and the overlap speech segments separately (Table \ref{tab:asr_wer_rA}). 
The proportion of single speaker segments (region-A) in the evaluation data is $24.8\%$.  
We see that the WER of the SCAM model is higher than the baseline systems for the single speaker regions (Region-A) but significantly better for the overlap speech regions (Region-B). Using clean embeddings improves the WER, suggesting  that the speaker activity conditioning in the proposed model helps in alleviating the problem of separating speakers while also providing accurate transcription for  overlapping speech. However, this is seen to come at the cost of slightly increased error rate in the single speaker regions over the BLSTM-iso system. 

\vspace{-5pt}
\section{Summary}\label{sec:conclusion}
A system for the transcription of natural conversations with multiple speakers is proposed in this paper. The speaker activity, predicted using a neural speaker diarization system, is used as the additional input to the acoustic model, to predict speaker specific senones in a  hybrid ASR system. The analysis of the proposed model shows implicit source separation and speaker specific embedding extraction  achieved in the proposed model. The advantage of the proposed model is also the ability to combine the speaker diarization and the acoustic model as a  single neural processing pipeline that can be jointly optimized. The experiments on the Switchboard dataset show the effectiveness of the proposed framework on two-speaker conversations in terms of significant improvements in the word error rates. 
\balance
\bibliographystyle{IEEEbib}
\bibliography{mybib}

\end{document}